\documentclass[11pt]{JHEP3}
\date{}

\usepackage{graphicx}
\usepackage{amssymb}
\usepackage{amsfonts}
\usepackage{epsfig}
\newcommand{\rdec}{r_\mathrm{dec}}
\newcommand{\rhor}{\rho_\mathrm{r}}
\newcommand{\fnl}{f_\mathrm{NL}}
\newcommand{\gnl}{g_\mathrm{NL}}
\newcommand{\Mp}{M_\mathrm{Pl}}

\title{The TeV-mass Curvaton}

\author{Kari Enqvist\\Physics Department and Helsinki Institute of Physics\\FI-00014 University of Helsinki, Finland\\ E-mail: \email{kari.enqvist@helsinki.fi}}
\author{Anupam Mazumdar\\ Physics Department, Lancaster University, Lancaster LA1 4YB, United Kingdom\\
Niels Bohr Institute, Copenhagen University, Blegdamsvej-17, Denmark }
\author{Olli Taanila\\ Helsinki Institute of Physics\\FI-00014 University of Helsinki, Finland\\ E-mail: \email{olli.taanila@iki.fi}}

\abstract{We consider the constraints for a curvaton with mass
$m\sim {\cal O}(1)$ TeV and show that they are not consistent
with a purely quadratic potential. Even if the curvaton
self-interactions were very weak, they must be accounted for as they
affect the dynamical evolution of the curvature perturbation. We
show that the only TeV-mass curvaton interaction potential that
yields the correct perturbation amplitude, decays before the
dark matter freeze-out, and does not give rise to non-Gaussian
perturbations that are in conflict with the present limits, is given
by $V_{\rm int}=\sigma^8/M^4$. The decay width of the curvaton
should be in the range $\Gamma=10^{-15}- 10^{-17}$ GeV.  The model typically predicts  large non-linearity parameters
$\fnl$ and $\gnl$ that should be observable by the Planck
satellite. We also discuss
various physical possibilities to obtain the required small curvaton decay rate.
 }

\preprint{HIP-2010-18}

\begin{document}

\section{\label{sec:introduction}Introduction}

In the curvaton mechanism \cite{curvaton}, the primordial
perturbations originate from quantum fluctuations of a light scalar
field $\sigma$ which during inflation gives a negligible contribution to the total energy
density (for a review, see~\cite{MR}). However, the
predictions of the curvaton model are quite sensitive to the form of
the curvaton potential and the dynamics before its decay. As discussed in \cite{EKM,dynamics,EKM1,kesn, kett,
subdominant,self-interacting}, even small deviations from
the extensively studied quadratic potential (see e.g. \cite{LUW}) can have significant
effects. This might appear paradoxical. However, one should bear in  mind that
the curvature perturbation $\zeta$ is a small number, which in the
$\Delta N$-formalism \cite{DeltaN} is the difference between the e-folds of two
separate FRW universes. Their expansion history is determined by the field dynamics,
which in the presence of even small non-linearities can lead to widely differing outcomes.
The number of e-folds in a given separate universe is a large number and its evolution
is smooth; however, the difference, which is a very small number, can be subject to
highly irregular oscillations, as was shown in \cite{subdominant}. The situation
is even more pronounced when discussing the non-Gaussianities, which are determined
by the field derivatives of the e-folds; here the non-linearity of the curvaton
potential is essential \cite{self-interacting}.

When decaying, the curvaton does not have to dominate the energy density of the universe.
It is enough that its perturbations are the dominant ones. However, in such a
subdominant curvaton case the decay products of the curvaton and the inflaton should
thermalize so as to convert the initial curvaton isocurvature perturbation into an
adiabatic one~\footnote{This is a real challenge for model building. Both the inflaton and the curvaton must decay into
the visible sector degrees of freedom. There are only few models where the curvaton carries the Standard Model charges;
a curvaton ~\cite{EKM,EKM1,mssmcv, AEM} based on the flat directions of MSSM (for a review, see~\cite{MSSM-REV}) is an example;
in few cases the MSSM curvaton may dominate the energy density while
decaying~\cite{mssmcv, AEM,john}.
}. In particular, the curvaton should decay before cold dark matter
particles freeze out; otherwise the CDM perturbation would not be adiabatic. Such
a requirement translates into a constraint on the decay rate $\Gamma$ of the curvaton. As
was pointed out already in \cite{subdominant,self-interacting}, the magnitude of $\Gamma$
is also essential for obtaining the observed amplitude $\zeta\sim 10^{-5}$: given some initial condition for the curvaton
field $\sigma$, for a fixed inflationary Hubble rate $H_*$, the curvaton has to decay
at a specific time in order to produce the correct perturbation.

%These are issues one should bear in mind when considering the interesting question of whether the curvaton could be a field accessible to present-day particle accelerators. A curvaton with a TeV mass scale could be some light modulus field, or it could be a flat direction of the MSSM (for reviews, see \cite{MSSM-REV}). MSSM curvatons \footnote{The MSSM Higgs can act as a curvaton in the (rather unlikely) case that the inflaton does not couple to the MSSM degrees of freedom at all \cite{mssmcv}.}, assuming curvaton dominance at the time of its decay, has been discussed in \cite{MSSMcurv} (see also \cite{john}), while the right-handed sneutrino as the curvaton candidate has been considered in i.e. \cite{RHNcurv}.

In the present paper we will first address the general question of whether a TeV mass scale curvaton
is possible at all, given the constraints on the curvature perturbation amplitude,
and on curvaton decay rate. In section \ref{importance} we demonstrate how in the specific case of a TeV mass curvaton
the free field assumption is not consistent with theoretical and observational constraints
and that in this case curvaton self-interactions cannot be neglected. In section \ref{selfintcurv} we scan numerically
the parameter space and compute the curvature perturbation, study different curvaton potentials and conclude that
only a curvaton with a self-interaction of the form $\propto \sigma^8$ is feasible. In section \ref{section:kinematic}
we discuss the origin of small curvaton decay width, $\Gamma\sim 10^{-17}-10^{-15}$ GeV. Although our treatment
assumes a real curvaton field, in the particle physics context (such as in MSSM) fields are typically complex,
and we point a possible mechanism for generating a small curvaton decay width that makes use of the kinematical
blocking due to the non-zero curvaton VEV.
Finally, section \ref{discuuss} contains a discussion of the results and possible
suggestions for future work.

\section{Importance of self-interactions}
\label{importance}

For a quadratic curvaton potential, both the field $\sigma$ and the perturbation $\delta\sigma$ have the same equation of motion \cite{LUW}. Hence the relative field perturbation stays constant until decay, $\delta\sigma/\sigma=\delta \sigma_*/\sigma_*$; here and throughout the paper we adopt the notation where $*$ denotes the initial value. Thus one can write the perturbation as
\[ \zeta \sim \frac{H_*}{\sigma_*} r_\mathrm{eff} \simeq 10^{-5}\; , \]
where $H_*/\sigma_*$ gives the initial perturbation amplitude in the curvaton, and $r_\mathrm{eff}$ is the efficiency factor that can be approximated quite well by the energy fraction at the curvaton decay \cite{LUW}:
\[ r_\mathrm{eff} \approx \rdec \equiv \left.\frac{\rho_\sigma}{\rhor+\rho_\sigma}\right|_{\hbox{decay}} \; .\]
Relating $\sigma_*$ and $r_*$ from $\frac{1}{2}m^2\sigma_*^2 / 3\Mp^2H_*^2 \simeq r_*$, and noting that $\rdec<1$, we find the constraint on the initial curvaton energy fraction
\begin{equation}
r_* <  \frac{1}{6}\frac{m^2}{\zeta^2 \Mp^2} \, . \label{eq:rzeta}
\end{equation}
In the free curvaton case $\rdec$ also determines non-Gaussianity through the simple relation \cite{LUW}
$\fnl = {5}/{4\rdec}$. Very roughly, observationally $|\fnl|<100$, which implies the constraint
\begin{equation}
r_* >  \frac{10^{-4}}{6}\frac{m^2}{\zeta^2 \Mp^2} \, . \label{eq:rnongauss}
\end{equation}

The limits (\ref{eq:rzeta}) and (\ref{eq:rnongauss}) are well known. However, there is more.
Since the observed perturbations are adiabatic to great accuracy, the curvaton must decay before dark matter decouples.
For each set of the initial conditions,
$(H_*, r_*)$, there is a relation between $\rdec$ and the effective decay constant $\Gamma$ given by
\begin{equation}
\rdec = \left.\frac{\rho_\sigma}{\rhor + \rho_\sigma}\right|_{H=\Gamma} \, . \label{eq:rdec_def}
\end{equation}
Here we assume implicitly a perturbative curvaton decay,
characterized by decay width $\Gamma$, but $\Gamma$ could stand for
any effective inverse decay time and thus the following discussion
should hold, at least roughly, also for a non-perturbative curvaton
decay as discussed in \cite{curvatondecres} (note however that non-perturbative curvaton decay
could turn out to be a source of a considerable non-Gaussianity \cite{Chambers:2009ki}).

The exact evolution of the energy densities is difficult to solve analytically. However, we can approximate the
curvaton evolution by dividing it up to three phases:
\begin{enumerate}
 \item When $V'' = m^2 < H^2$, the curvaton is effectively massless, so the field value stays constant, $\sigma = \sigma_*$.
 \item When $V'' = m^2 > H^2$ the curvaton oscillates in the quadratic potential, and thus its energy density approximately scales as $\rho_\sigma \propto a^{-3}$ \cite{Turner:1983he}.
 \item The curvaton oscillates until $H = \Gamma$, whence it decays.
\end{enumerate}
Solving the Friedmann equation for the regime where $m^2 > H^2$ then yields
\[ \frac{a(H)}{a_*} = \sqrt{\frac{H_*}{H}} \left\{ 1 + \frac{r_*}{4} \left[ \frac{H_*^2}{m\sqrt{Hm}}-1\right]\right\} + \mathcal{O}\left( r_*^2\right)
\; .\]
Using the above result we can solve eq.~(\ref{eq:rdec_def}) to give
\begin{equation}
r_* = \frac{m\sqrt{m \Gamma}}{H_*^2} \frac{6 \left( \frac{\Mp}{m}\right)^2 \zeta^2}{\frac{H_*^2}{m\sqrt{m\Gamma}} - 12 \left( \frac{\Mp}{m}\right)^2\zeta^2} \; . \label{eq:rgamma}
\end{equation}
To make sure that the curvaton decays before dark matter decouples, $\Gamma$ needs to be large enough. For, i.e., a LSP originating from MSSM with
a decoupling temperature $T_{\rm LSP}\sim {\cal O}(10)$ GeV, one would obtain as a conservative limit
$\Gamma \geq 10^{-17} \, \mathrm{GeV}$, which we will adopt as a benchmark value for the rest of the paper.

\EPSFIGURE[t]{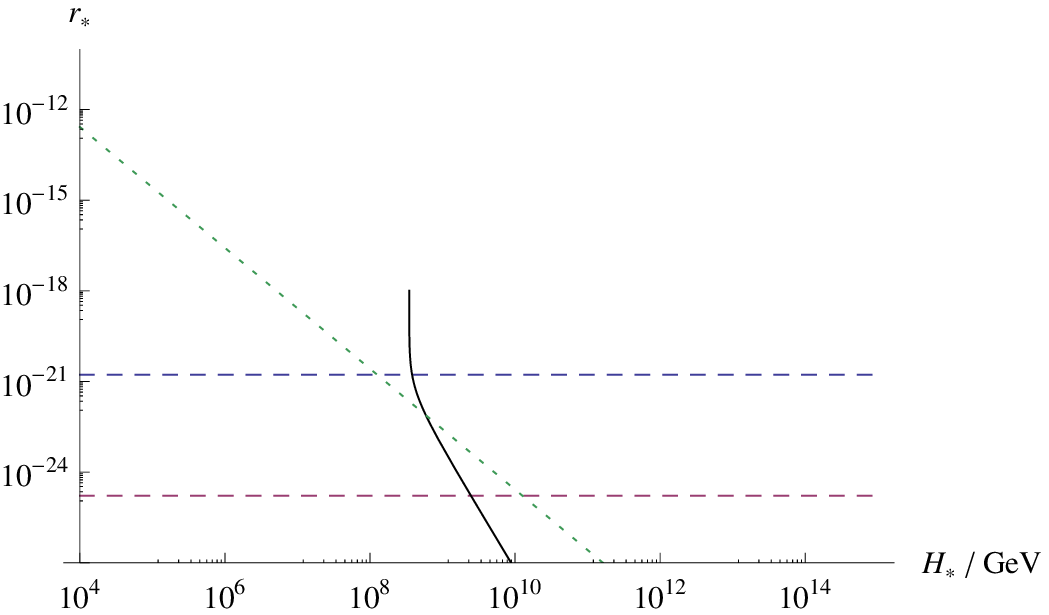,width=12cm}{\label{fig:quadratic}Parameter space of the quadratic curvaton. $r_*$ must be above the red dashed line to produce $\zeta \sim 10^{-5}$ (equation (\ref{eq:rzeta})) and below the blue dashed line to produce small enough $\fnl$ (equation (\ref{eq:rnongauss})). Furthermore, $\Gamma$ is constrained from above, and thus only the parameter space to the right of the black solid line is allowed (equation~(\ref{eq:rgamma})). The green dotted line illustrates the equality of the mass term and a possible self-interaction term in the potential (equation (\ref{eq:req})) for $n=4$. For smaller values of $n$ the line moves further to the left. To the right of the dotted line the self-interaction dominates, and thus practically in all of the allowed parameter space the self-interaction must be taken into account. }

In Fig.~\ref{fig:quadratic} we have plotted the conditions (\ref{eq:rzeta}), (\ref{eq:rnongauss}) and (\ref{eq:rgamma}) for $m = 1\,\mathrm{TeV}$; the first two
appear as two horisontal dashed lines, and $r_*$ is limited to be between those two. The condition (\ref{eq:rgamma}) appears as black
solid line and the allowed region is to the right. Thus, na\"{i}vely, there would appear to be a large
parameter region
with $r_* \sim  10^{-25}\ldots10^{-21}$ and $H_* \gtrsim 10^9 \, \mathrm{GeV}$ where a TeV-mass curvaton is allowed.

However, since the curvaton must decay, it has to have interactions; integrating these out would result in an effective theory with some self-interactions. Hence the curvaton potential cannot be purely quadratic,
although the self-interaction can be very weak, either in the sense that it is a Planck scale suppressed,
or alternatively, if the relevant mass scale is lower, the corresponding coupling constant is very small. Let us therefore write
\begin{equation}
V(\sigma) = \frac{1}{2} m^2 \sigma^2 + \frac{\sigma^{n+4}}{\Mp^n} \; . \label{eqn:potential}
\end{equation}
In order for the quadratic assumption to be consistent,
we need to require that
\[ \frac{1}{2}m^2 \sigma^2 \gg \frac{\sigma^{n+4}}{\Mp^n}\]
throughout the evolution. Since the energy density of the quadratic field decreases monoton\-ously, it is sufficient to apply this requirement
only for the initial conditions.
Solving for $r_*$ such that the magnitudes of the quadratic and non-quadratic terms are equal, we find the condition
\begin{equation}
 \label{eq:req}
r_* = \frac{m^2}{3\Mp^2H_*^2} \left( \frac{m^2 \Mp^n}{2} \right)^\frac{2}{n+2} \; .
\end{equation}
We have plotted this condition for $n=4$ in figure \ref{fig:quadratic} as the green dotted line. To the right of it, the non-quadratic term
dominates initially.
As can be seen in figure \ref{fig:quadratic}, there is practically no allowed region in the parameter space where the quadratic assumption
would even approximately apply. For smaller values of $n$,
the self-interaction becomes important even for much smaller values of $H_*$ and $r_*$, and thus, there is no quadratic regime left in the
parameter space.

We thus conclude that even if the curvaton self-interactions are very weak,
a purely quadratic potential would not be a consistent approximation for a mass $m\simeq 1\,\mathrm{TeV}$;
instead, the effects of the self-interactions need to be taken into account. As we will next discuss,
the self-interactions change the dynamics of the curvaton in a significant way.

\section{The self-interacting TeV mass curvaton}
\label{selfintcurv}

Let us now study the dynamical evolution of the perturbations in the curvaton potential (\ref{eqn:potential}) in more detail.
We calculate the amplitude of perturbations and the non-Gaussianity parameters using $\Delta N$-formalism~\cite{DeltaN}.
This equals to solving the set of equations
\begin{eqnarray}
 0&=&\ddot{\sigma}+\left(3H + \Gamma\right)\dot{\sigma} + m^2 \sigma + (n+4)\frac{\sigma^{n+3}}{\Mp^n}  \label{eqn:eom1}\\
\dot{\rhor} &=& -4H\rhor + \Gamma \dot{\sigma}^2\label{eqn:eom2}\\
3H^2\Mp^2 &=& \rhor + \frac{1}{2}\dot{\sigma}^2 + V(\sigma) \quad , \label{eqn:eom3}
\end{eqnarray}
in two different, causally disconnected patches of the universe. Here we have assumed that the curvaton decay can be at least qualitatively described by an effective decay constant
$\Gamma$ and that ultimately the curvaton decays to radiation $\rhor$.

The initial conditions are given by $\rhor = 3\Mp^2H_*^2$ and $\dot{\sigma} = 0$. However, the initial value of the curvaton is not equal in the two patches. Instead, we assume the values $\sigma_1 = \sigma_*$ and $\sigma_2 = \sigma_* + H_*/2\pi$, where $V(\sigma_*) = r_* \rhor$. The curvature perturbation is then given by the difference in e-folds, $\zeta = \Delta N = N(\sigma_* + H_*/2\pi) - N(\sigma_*)$, where $N$ is the number of e-folds evaluated at a fixed $H$ after the curvaton has decayed. We then scan through the space of all possible initial conditions $(H_*, r_*)$, and at each point adjust $\Gamma$ so that the right amplitude of the perturbations is achieved. We then calculate $\fnl$ and $\gnl$ for each point, given by the expressions
\[ \fnl = \frac{5}{6} \frac{N''}{N'^2} \qquad \mathrm{and} \qquad \gnl = \frac{25}{54} \frac{N'''}{N'^3} \; ,\]
where the prime denotes a derivative with respect to the initial condition. This results in an allowed region of the parameter space that is limited by the following factors:
\begin{enumerate}
 \item The initial perturbation in the curvaton field is too small, i.e., the ratio $H_*/\sigma_*$ is so small, that even if the curvaton is completely dominant,
 the final perturbations are nevertheless too small.
 \item The isocurvature limit: To make sure that the curvaton decays only to adiabatic modes, we must require that it decays before the dark matter particles decouple from radiation background. As before,
 we take this to translate into the requirement $\Gamma \geq 10^{-17} \, \hbox{GeV}$.
 \item The non-Gaussianity parameters $\fnl$ and $\gnl$ must be within the observed limits. Here we use the same limits as in \cite{self-interacting}, which are given by $-9 < \fnl < 111$ \cite{fnllimit} and $-3.5 \times 10^5 < \gnl < 8.2 \times 10^5$ \cite{gnllimit}.
\end{enumerate}
The first condition limits the allowed parameter space mainly from upper left, the isocurvature limit from lower left, and the non-Gaussianity limit from lower right.

\subsection{Different powers of the self-interaction}

\EPSFIGURE[t]{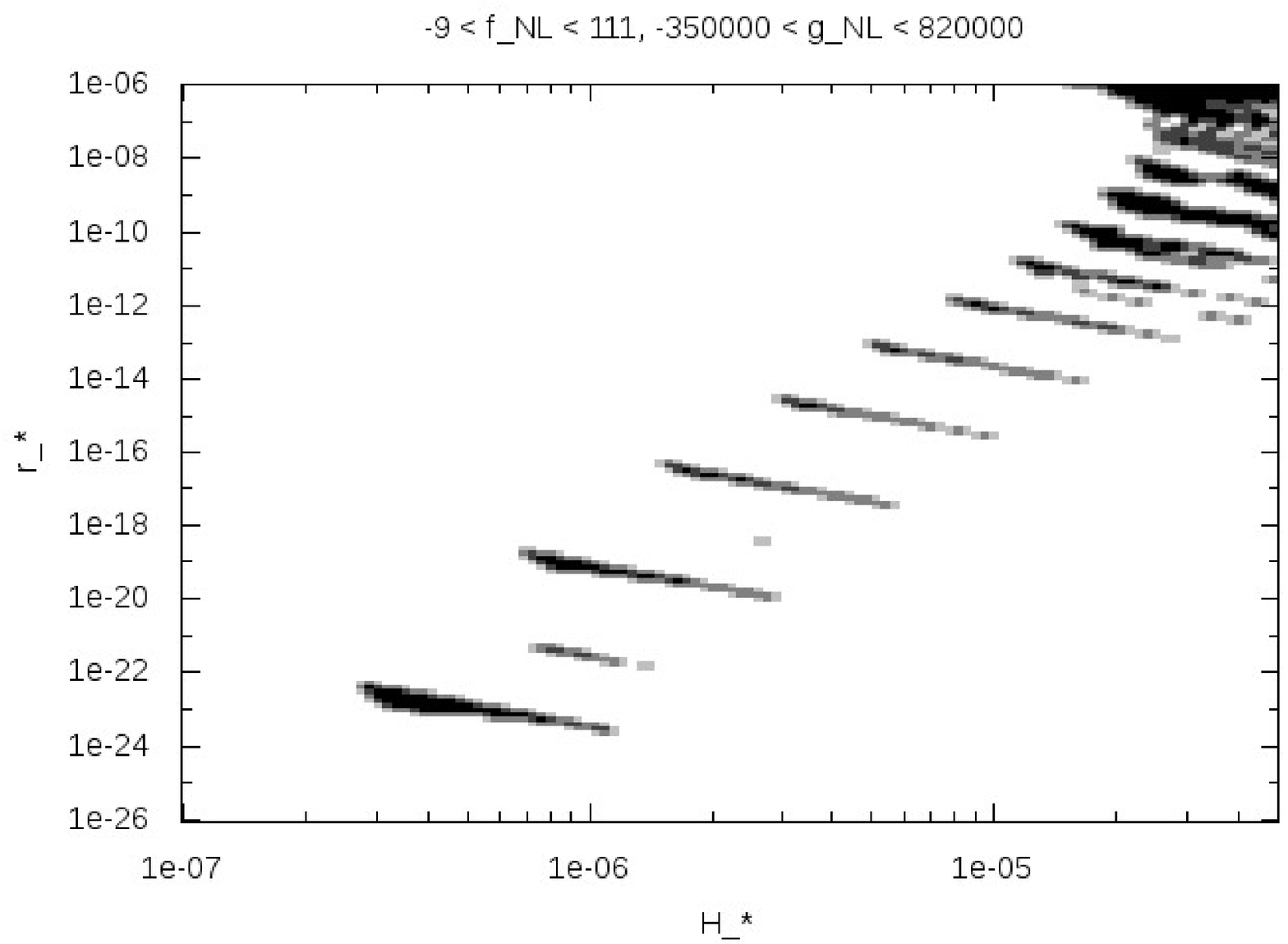,width=10cm}{\label{fig:fnlgnl}The allowed region in the parameter space for $n=4$.
Since the values of $\Gamma$ and $\rdec$ oscillate in the
non-quadratic regime, so do $\fnl$ and $\gnl$. Thus the requirement for the non-Gaussianity parameters creates the
distinctive striping. Note that the units of $H_*$ are $\Mp$.}

The value of $n$ is \emph{a priori} unconstrained. However, from \cite{subdominant} we know, that if $n\geq6$, the field evolves smoothly also when in the non-quadratic
part of the potential, and begins oscillations only when it enters the quadratic regime. Due to the lack of oscillations in the non-quadratic regime, there are no enhancement of the perturbation or the non-Gaussianity parameters. Indeed, we have scanned through the parameter space for $n\geq6$ numerically and found no allowed parameter space left.

A relatively similar case is $n=0$. Here the field oscillates in the non-quadratic regime,
but these oscillations are so fast that the behaviour in parameter space is still smooth.
We have scanned the parameter space numerically to confirm that for $n=0$, there is no allowed parameter space left.

Thus we are left with $2 \leq n \leq 6$. As was demonstrated in \cite{subdominant} there are strong oscillations in the parameter space,
and thus one can expect enhancement of $\zeta$ in the non-quadratic regime for these choices of $n$. However, after performing a
systematic scan through the parameter space, we are left with the surprising conclusion that $n=2$ is disallowed for a
$1 \, \mathrm{TeV}$ mass. For $n=4$ there is however still some parameter space left. These are illustrated as the black points in
figure \ref{fig:fnlgnl}.

We thus come to the following conclusion: A curvaton with a $1 \, \mathrm{TeV}$ mass, and with a potential of the form
\[ V(\sigma) = \frac{1}{2}m^2\sigma^2 + \frac{\sigma^{n+4}}{\Mp}\]
can produce the observed primordial perturbations only if $n=4$.

\subsection{Predictions for $n=4$}

The black pixels in figure \ref{fig:fnlgnl} correspond to those initial conditions $(H_*, r_*)$,
which produce $\zeta \sim 10^{-5}$, while still decaying early enough to avoid isocurvature and producing
non-Gaussianity which is within the observational limits. The values of $\fnl$ and $\gnl$ change widely from pixel to pixel,
due to the oscillations present in the non-quadratic regime. In particular, they will change sign;
hence there will always be regions in the parameter space where they are sufficiently small.
However, the average absolute value of both $\fnl$ and $\gnl$ is so large, that the values of $\fnl$
and $\gnl$ in the allowed regions are, except for some singular points, typically detectable with future
experiments, i.e., the Planck satellite. This is demonstrated in figure
\ref{fig:nongauss} for a fixed value of $H_* = 10^{12} \, \mathrm{GeV}$.
%The oscillations make the allowed patches isolated from each other, and thus the allowed parameter space can be characterized as a group of diagonal stripes.

\EPSFIGURE[t]{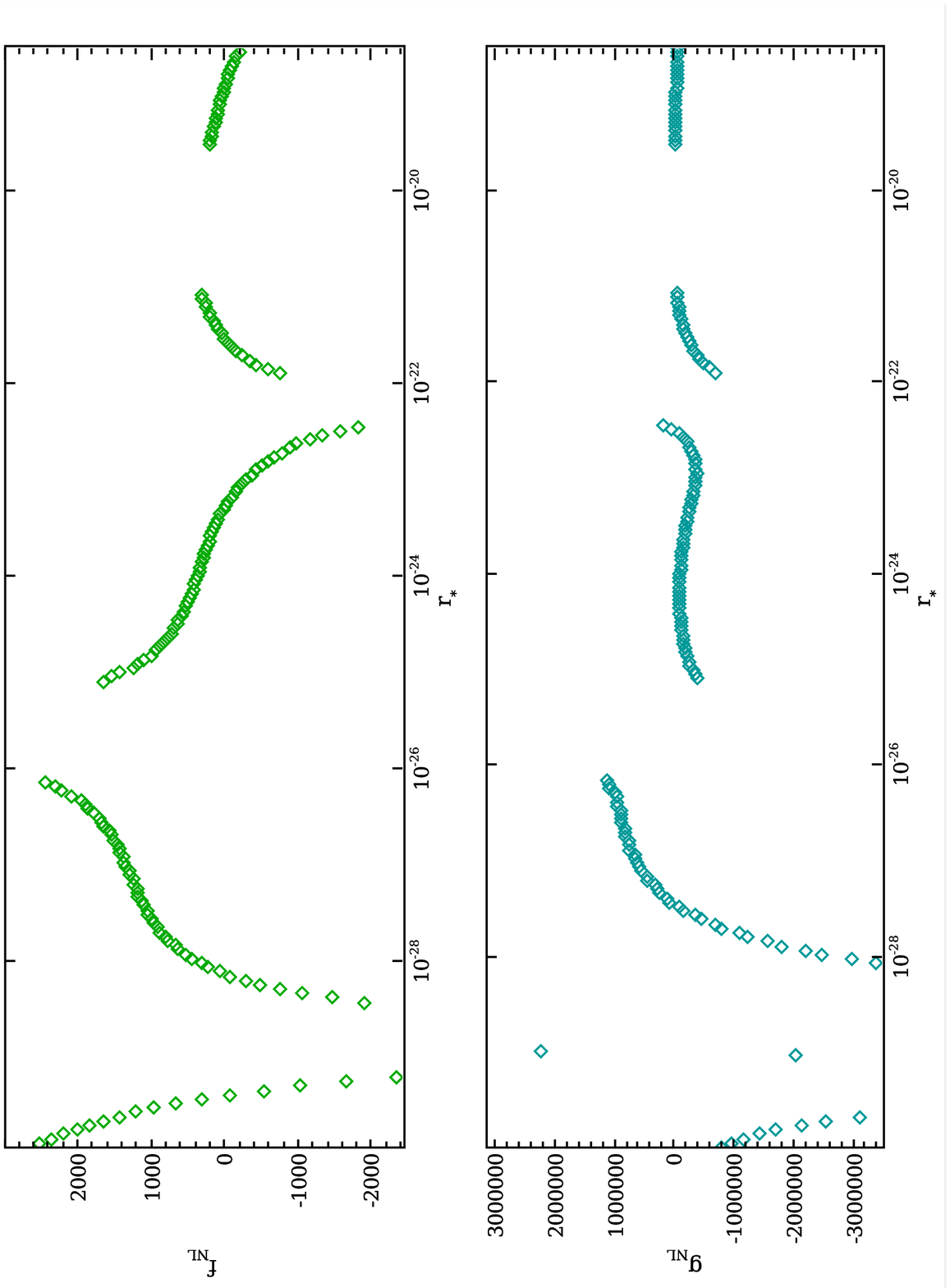,width=10cm,angle=270}{\label{fig:nongauss} $\fnl$ and $\gnl$ plotted against $r_*$ for all points producing $\zeta \sim 10^{-5}$ and for a fixed $H_* = 10^{12}\,\mathrm{GeV}$.}

In figure \ref{fig:gamma} the corresponding values for $\Gamma$ are given, which demonstrates that the allowed range is given roughly by $10^{-15}\ldots10^{-17} \, \hbox{GeV}$, i.e., the allowed points mostly populate the lower limit for $\Gamma$.

In general, the numerical results can be summarised to be the following:
\begin{itemize}
 \item A curvaton with a mass of $1 \, \mathrm{TeV}$ is limited to the region in the parameter space where self-interactions suppressed by the Planck mass become important.
 \item Of all the possible powers of the self-interactions, only the case $n=4$ has non-negligible allowed parameter space. This corresponds to a self-interaction of the
 type $V_\mathrm{int}\propto \sigma^8$.
 \item The typical values of both $\fnl$ and $\gnl$ are so large in the allowed patches of the parameter space, that they are observable with future CMB observations.
\end{itemize}

\EPSFIGURE[t]{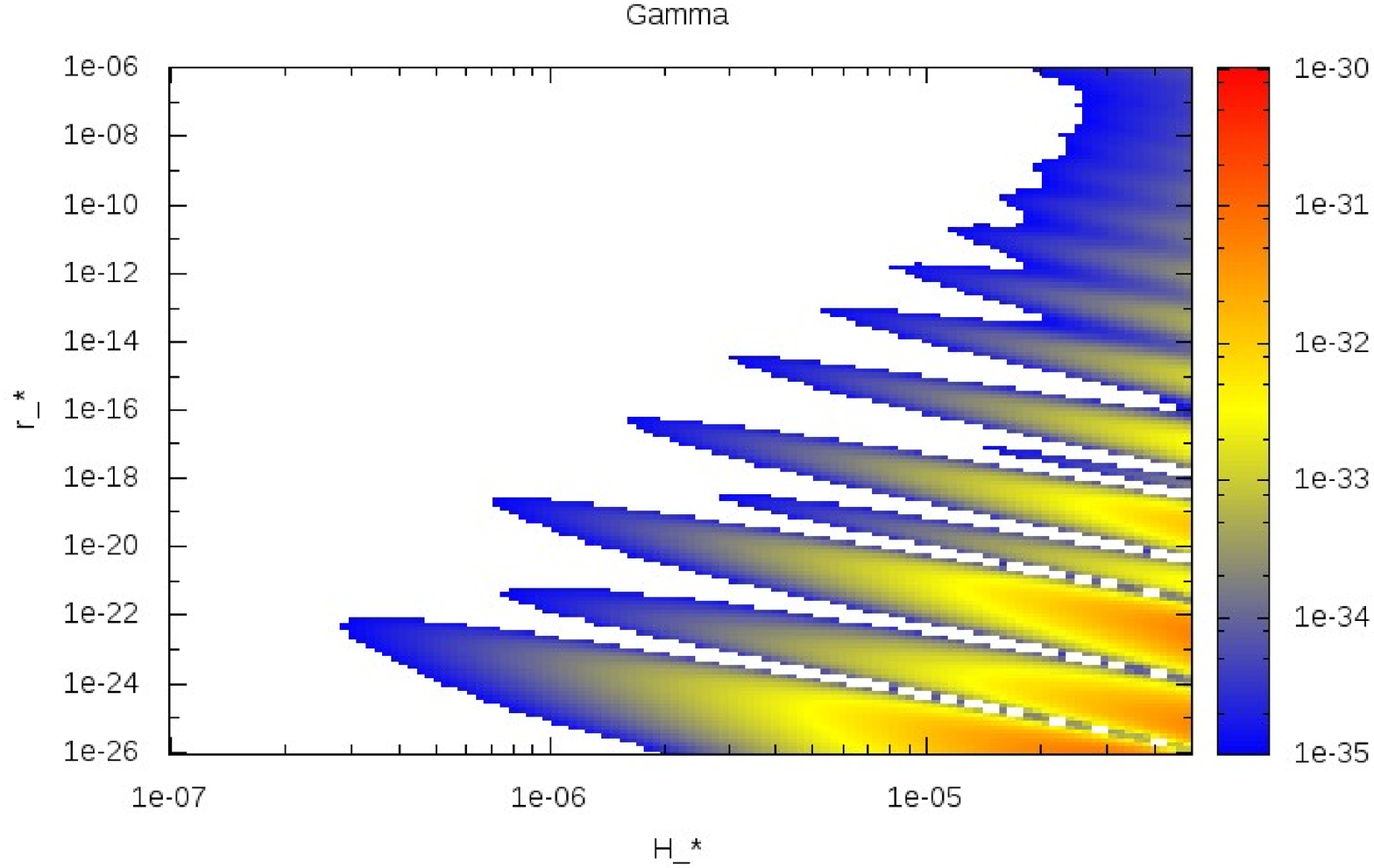,width=10cm}{\label{fig:gamma}The values of $\Gamma$ corresponding to the $\zeta \sim 10^{-5}$ for $n=4$. The units of $H_*$ and $\Gamma$ are in $\Mp$. Note that in comparison to figure~\ref{fig:fnlgnl}, points with too large $\fnl$ or $\gnl$ have not been removed from this plot. }

Even though we have here solved the equations numerically for mass of $1 \, \mathrm{TeV}$, we can deduce the results also for other masses. From the analysis of \cite{subdominant} and \cite{self-interacting} it is clear that enlarging the mass will also enlarge the allowed regions of the parameters space. Thus for a $10 \, \mathrm{TeV}$ mass, the plot of the allowed points in the parameter space would look otherwise similar to figure \ref{fig:fnlgnl}, but the black stripes would be somewhat thicker.
Similarly, for smaller masses, i.e.~$100~\mathrm{GeV}$, the allowed region will become smaller.
For a $10~\mathrm{GeV}$ mass practically no parameter space would be left.

\section{Source of small $\Gamma$: Kinematical blocking}
\label{section:kinematic}

As demonstrated by figure \ref{fig:gamma}, the allowed values for the decay constant $\Gamma$ populate the lower bound, i.e., are in the range $10^{-15}\ldots 10^{-17} \, \mathrm{GeV}$. Although we have not constrained ourselves to any single particle physics model, for most realistic scenarios this coupling is very small. To alleviate this problem, we suggest possible mechanism to implement such a small decay constant.

After inflation the majority of the energy density of the universe is in the relativistic degrees of freedom. We strictly assume that they belong to the
visible sector degrees of freedom which would thermalize with that of the Standard Model quarks and leptons. Those degrees of freedom which are coupled to the curvaton will obtain VEV dependent masses, which for a quartic coupling $V_\mathrm{int} \sim h^2 \sigma^2 \phi^2$ is $m_\mathrm{eff} \sim h \langle \sigma \rangle$. The decay of the curvaton is thus delayed until $m_\mathrm{eff} < m\sim  {\cal O}(1)$~TeV. To estimate when the curvaton decays, one has to follow the evolution of the VEV of the curvaton to find out when $m \sim h\langle \sigma\rangle$.

For a real valued curvaton, the curvaton field oscillates through zero, and thus the kinematical blocking is lifted for a short part during each oscillation. In such a case, however, instant preheating-type phenomena will typically become important~\cite{Instant}, see for a review~\cite{PR-REV}.
For a complex field the field trajectory cannot pass the origin if there is any initial rotational motion~\cite{AM4}.
Strictly speaking, our analysis in the preceding section is valid for real fields alone. However, starting from the action for a complex field,
\[ \mathcal{L} = \frac{1}{2}\partial^\mu \varphi \partial_\mu \varphi^* - \frac{1}{2}m^2|\varphi|^2 - \frac{|\varphi|^{n+4}}{\Mp^n} \; ,\]
and inserting an Ansatz $\varphi = |\phi| e^{i\theta}$
yields an effective action of the real modulus field with an effective mass
\[m_\mathrm{eff}^2 = m^2 - \dot{\theta}^2 \, .\]
Hence we may still adopt our analysis in the preceding sections for the real field if the rotational motion is not too fast. Equivalently one could write the equations of motion for the field $\phi \rightarrow \phi e^{i\theta}$ (without the Hubble friction term) as
\begin{eqnarray*}
\frac{d}{dt}\left( \dot{\theta}\phi^2\right) &=& 0\\
\ddot{\phi} + m^2 \phi - \frac{L^2}{\phi^3} &=& 0 \; ,
\end{eqnarray*}
where $L$ is a constant associated with angular momentum. The evolution of the radial component of the field is similar to the real field case for large values of
$\phi$, but near the origin the effective potential rises fast, and the radial part is reflected away from the origin. If the time that the field spends near the origin is
very small compared to the background dynamics (Hubble time), then the evolution of the complex field will effectively be the same as for the real field. This requires small enough $L$.

Thus for the decay of a complex field to be kinematically blocked, we require that there is sufficient rotational motion as to prevent the field from traversing too close to the origin, yet at the same time to be small enough for the analysis of the preceding sections to apply. Here we assume that these conditions are met.

In the quadratic case, the VEV of the curvaton will be redshifted as $a^{-3/2}(t)$ once the flat direction commences its oscillations at $H(t)\sim m$.
In a radiation dominated epoch the scale factor is $a(t)\propto H^{-1/2}(t)$. Thus 
\begin{equation}
 \label{eq:sigmaH} \sigma(H) = \sigma_*\left( \frac{H}{m} \right)^\frac{3}{4} \; .
\end{equation}
Relating the initial field value $\sigma_*$ to $r_*$ used in figures \ref{fig:fnlgnl} and \ref{fig:gamma} with the equation $r_* = \frac{1}{2}m^2\sigma_*^2 / 3\Mp^2H_*^2$, we can solve the value of $H$ at the moment the curvaton starts to decay:
\[ H_\mathrm{dec} = m \left( \frac{1}{\sqrt{6r_*}h \Mp H_*} \right)^\frac{4}{3}\]

In contrast, in the non-quadratic case, the VEV starts to decrease when the curvaton becomes massive,
$H < H_\mathrm{sr} \equiv \sqrt{m^2+56\,\sigma_*^6/\Mp^4}$, while the field scales as $\sigma \propto a^{-3/5}$, and
\begin{equation}
\label{eq:non-quadratic}
\sigma(H) = \sigma_*\left( \frac{H}{H_\mathrm{sr}}\right)^\frac{3}{10} \; .
\end{equation}
This continues until the VEV has diminished sufficiently and enters into the quadratic regime. The point of transition is defined by
$\frac{1}{2}m^2\sigma_\mathrm{eq}^2 \approx \sigma_\mathrm{eq}^8/\Mp^4$. After this point the VEV behaves as in the quadratic case with
\[ \sigma(H) = \sigma_\mathrm{eq} \left( \frac{H}{H_\mathrm{eq}} \right)^\frac{3}{4} \; . \]
Now again requiring that the curvaton decays when $h\sigma \approx  m$, we get
\begin{equation}
\frac{m}{h} = \sigma(H_\mathrm{dec}) = \sigma_\mathrm{eq} \left( \frac{H_\mathrm{dec}}{H_\mathrm{eq}} \right)^\frac{3}{4} \;
\end{equation}
Solving $H_\mathrm{eq}$ from Eq.~(\ref{eq:non-quadratic}),
\begin{eqnarray}
H_\mathrm{eq} = H_\mathrm{sr} \left( \frac{\sigma_\mathrm{eq}}{\sigma_*} \right)^\frac{10}{3}\,,~~~~~~
%we can finally write
 H_\mathrm{dec} = H_\mathrm{eq}\left( \frac{m}{h \sigma_\mathrm{eq}} \right)^\frac{4}{3}\,.
\end{eqnarray}
Inserting the expressions for $H_\mathrm{eq}$ and $\sigma_\mathrm{eq}$, we get
\begin{eqnarray}
H_\mathrm{dec} &=& H_\mathrm{sr}\left( \frac{\sigma_\mathrm{eq}}{\sigma_*} \right)^\frac{10}{3}\left(\frac{m}{h\sigma_\mathrm{eq}}\right)^\frac{4}{3}\nonumber\\
&=& \sqrt{m^2+\frac{56\sigma_*^6}{\Mp^4}}\left[\frac{ \left( \frac{m^2\Mp^4}{2} \right)^\frac{1}{6}}{\sigma_*}\right]^\frac{10}{3}
\left[\frac{m}{h\left( \frac{m^2\Mp^4}{2} \right)^\frac{1}{6}}\right]^\frac{4}{3} \;\label{eqnabove} .
 \end{eqnarray}
Solving $\sigma_*$ from $r_* = \sigma_*^8 /\Mp^4 \, / \, 3\Mp^2 H_*^2$, and inserting it into (\ref{eqnabove}), we get
\begin{equation}
H_\mathrm{dec} = \frac{m^2 \sqrt{m^2+56\,H_*^\frac{3}{2}r_*^\frac{3}{4}\Mp^\frac{1}{2}}}{2^\frac{1}{3}3^\frac{5}{12}
\Mp^\frac{7}{6}h^\frac{4}{3}r_*^\frac{5}{12}H_*^\frac{5}{6}} \; .
\end{equation}

\EPSFIGURE[t]{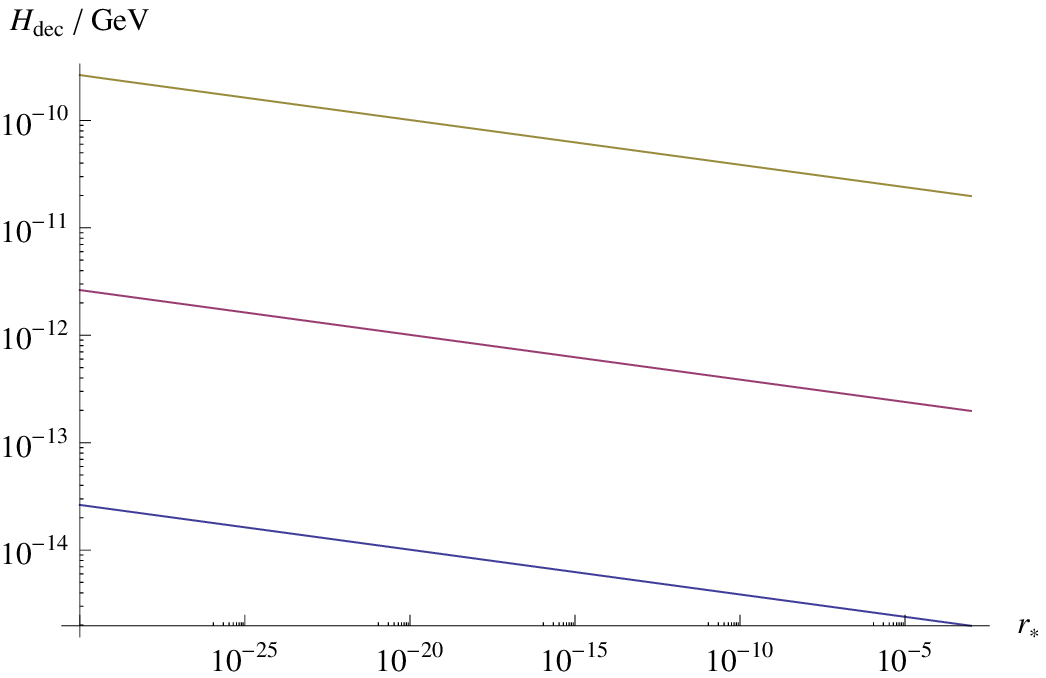,width=10cm}{\label{fig:decay}Value of $H_\mathrm{dec}$ plotted for $H_* = 10^{13} \; \mathrm{GeV}$ and
for masses $1000$, $100$ and $10 \; \mathrm{GeV}$.}

%Assuming that the flat direction decays when the effective mass of the decay products is smaller than the mass of the flat direction, i.e., $h \sigma < m$, the decay happens when the Hubble expansion rate is given by:

The curvaton decay takes place when $\Gamma\sim H_{dec}$.  The perturbative decay rate is determined by $\Gamma_d\sim h^2m$, which is larger than the kinematical blocking, i.e. $\Gamma_d\gg \Gamma\sim H_{dec}$. Therefore once the kinematical blocking is lifted, the curvaton decays immediately~\cite{AM1,AM2,AM4,PR-REV}.

The values of $H_\mathrm{dec} \sim \Gamma$ for some typical values of $H_*$ and $r_*$ are given in figure \ref{fig:decay}.
The time of decay is relatively insensitive to all the other parameters, and strongly decreases as the function of the curvaton mass.
A relatively small $\Gamma \sim 10^{-12}~\mathrm{GeV}$ can be obtained for small masses (i.e. $m\simeq 100~\mathrm{GeV}$)
for a coupling of order unity, $h\sim {\cal O}(1)$.  The perturbative value of $\Gamma_d$ is generically quite large for $h\sim {\cal O}(1)$. To acccommodate the small value for $\Gamma \sim 10^{-15} -10^{-17} \, \mathrm{GeV}$, the curvaton must couple to the SM degrees of freedom
very weakly, i.e. $h\sim 10^{-9}-10^{-10}$. Unfortunately, within MSSM it is not possible to find such a
curvaton candidate unless the decay of the curvaton is somehow delayed.

One such attractive possibility may arise if the curvaton instead of decaying directly, first fragments into lumps of
supersymmetric matter, known as Q-balls~\cite{Coleman}. The formation of Q-balls within MSSM
has been extensively studied analytically~\cite{Kusenko} and numerically~\cite{Kasuya}.
The Q-ball formation is quite robust even if the curvaton oscillates radially~\cite{ESM},
all that we require is that the curvaton oscillations on average feel negative pressure.
In presence of a negative pressure certain sub-Hubble modes of the curvaton perturbations become unstable and leads to the
formation of Q-balls. These Q-balls do not decay immediately, instead they evaporate into pair of fermions from their
surface~\cite{Coleman}, naturally suppressing the effective decay rate~\cite{ESM} (for a review, see~\cite{MSSM-REV}).
However the details of the decay rate depends on the charge accumulated in the
Q-balls~\footnote{One of the astrophysical signature of fragmentation of any condensate is the
production of very high amplitude gravity waves which can be detectable by future gravity wave detectors~\cite{KM} .}.

There are three possibilities emerge within MSSM which can fragment to form Q-balls.
They are the $udd$, $LLe$ and $QLd$ monomials~\cite{MSSM-REV}. Some of these monomials are
lifted by non-renormalizable superpotential operators of dimension $d=5$, such as $H_uLLLe,~H_uLudd$.
Typically these superpotential operators are hybrid in nature and would never allow a stable non-renormalizable
$A$-term in the potential proportional to the superpotential itself. As  a result the potential for such flat
directions would look exactly like Eq.~(\ref{eqn:potential}) with $n=4$.

%It should be noted that the calculation presented here is only an approximate, and a detailed (numerical) study might give results differing by even orders of magintude to either direction.

%\subsection{Q-balls}

%Another alternative is to consider a scalar field which does not decay into particles directly. For example some MSSM flat directions decay into \emph{Q-balls}

%\section{TeV-mass curvaton and MSSM}
%\label{andMSSM}

%Note that  all the relativistic species of MSSM are not excited right after the inflaton decay. Those MSSM degrees of freedom which are coupled to the flat direction will obtain VEV dependent masses and would slow down the rate of thermalization which are mediated by the gauge bosons and gauginos. Now in the kinematic blocking calculation, the effective mass is given by $m_{eff}\sim h\sigma(t)$, where $h$ denotes a Yukawa or gauge coupling coupling, and applies to those scalar which are not on the D-flat sub-space, and their fermionic partners, through the F-terms.

%The perturbative value of $\Gamma$ for these flat directions are however far too large. To acccommodate the small value fo $\Gamma \sim 10^{-15} \ldots 10^{-17} \, \mathrm{GeV}$, the scenario must be augmented with i.e.\ kinematic blocking. The na\"{i}ve estimate of section \ref{section:kinematic} however gives $h \sim \mathcal{O}(1)$ from the top-quark Yukawa-coupling, and thus for $m = 1 \, \mathrm{TeV}$ this gives $\Gamma$ which is too large by a couple of orders of magnitude. \textbf{Here should be additional babble.}

\section{Discussion}
\label{discuuss}

In this paper we have considered requisites for the existence of a TeV mass curvaton. Such a curvaton has to produce the observed perturbation amplitude and not too much non-Gaussianity; moreover, it has to decay before the CDM freeze-out. These constraints fix the range of the initial conditions which turns out to be such that the quadratic term in the curvaton potential cannot dominate over possible higher-order terms for the whole dynamical range. Hence we conclude that a purely quadratic curvaton potential would not be a consistent description of a TeV mass curvaton. However,  the presence of non-linearities in the field equation of motion then very much modifies the outcome for value of the curvature perturbation. Interestingly enough, as discussed in Sect.~\ref{selfintcurv}, we find that the only viable curvaton potential that satisfies all the observational constraints is $V=m^2\sigma^2/2+\sigma^8/M^4$. Moreover, the curvaton decay rate should be in the range $\Gamma=10^{-15}- 10^{-17}$ GeV. Note that in the case where the curvaton energy density is subdominant at the time of decay, the curvaton does not necessarily have to decay before baryogenesis, which can be a process that takes place among the inflaton decay products. However, the decay should be able to produce thermal CDM particles so that the CDM perturbation is adiabatic.

As we point out, MSSM flat directions which are lifted by $d=5$ operators would produce exactly the form of the potential required for the TeV mass curvaton. Here we should emphasize the fact there are directions with no $A$-terms so that although the flat direction field is complex, its modulus would, to a good approximation, have the same equation of motion as the real field considered in  Sect.~\ref{selfintcurv}. Rotational motion in the complex field plane would effectively only modify value of the mass parameter, and as long as the angular momentum is small enough, the results of Sect.~\ref{selfintcurv} should apply. However, the rotational motion is essential for the kinematical blocking that protects the flat direction field from decaying as long as
the field is away from the origin.

As discussed in Sect.~\ref{section:kinematic}, kinematical blocking is also modified the presence of non-linearites. We find an effective $\Gamma$ that is small, but perhaps not small enough. For some regions in the parameter space, especially for small curvaton mass parameters, the correct perturbation can nevertheless be generated. We should however stress the approximate nature of our estimate. A proper calculation would account not only for the actual rotational motion but also the dynamics (and possible backreaction) of the
kinematically blocked degrees of freedom. This would provide a major numerical challenge but might nevertheless be worth the attempt. Meanwhile, we can only conclude that a TeV mass curvaton based on a $d=5$ MSSM flat direction remains a possibility. Another natural solution may arise if the flat direction fragments to form Q-balls, which then evaporate via surface evaporation which can delay the decay rate sufficiently.

Note also that what really matters is the equation of state, not the time of decay.
Thus if the curvaton decays too early, the perturbations might still generated if the
decay products have the equation of state of matter. An example of this could be the MSSM flat direction decaying into Q-balls, which would then slowly decay. Another possibility could be that the MSSM curvaton would not be reponsible for the amplitude of the perturbations, but might add a small component on top of the inflaton perturbation spectra
with (very) large values of $\fnl$ and $\gnl$. The production of large non-Gaussianities by the MSSM flat directions could in fact be a generic feature that would deserve a closer study.

\acknowledgments
This research is supported by the
European Union through Marie Curie Research and Training Network
``UNIVERSENET'' (MRTN-CT-2006-035863). KE is also supported by the Academy of Finland
grants 218322 and 131454.
O.T. is supported by the Magnus Ehrnrooth foundation.

\end{document}